\documentclass[smallextended]{svjour3}       
\usepackage{graphicx}
\usepackage{longtable}
\usepackage{graphicx}
\usepackage{epstopdf}
\usepackage{amsmath,amssymb,amsfonts}
\usepackage[usenames]{color}
\usepackage{url}
\usepackage{graphicx,tikz,pgf,epsf}
\journalname{Arxiv}
\begin{document}

\title{Quantum Teleportation And Tripartite Quantum State Sharing using a Seven Qubit Genuinely Entangled State
}
\author{Mrittunjoy Guha Majumdar	\and
		          	Prasanta Panigrahi 
}

\institute{M. Guha Majumdar \at
		St. Stephen's College, Delhi, India\\ \\
		    \and
            P. Panigrahi \at
		Department of Physics, 
		IISER, Mohanpur, India
}

\date{Accepted: 16 December 2014}

\maketitle

\begin{abstract}
We have tried to gauge the utility of a recently introduced seven-qubit state by Xin-Wei Zha et al for quantum teleportation and quantum state sharing. It is shown that this state can be used for the teleportation of an arbitrary single, double and triple qubit systems. We have also come up with 3 proposals for Quantum State Sharing using the Xin-Wei Zha state. 

\end{abstract}

\section{Introduction}
Quantum Entanglement has been one of the key ideas in the revolution that the contemporary world has witnessed with the emergence of the concept of Quantum Computation and Information. It has been used to formulate certain interesting, though often counterintuitive, applications such as teleportation and super-dense coding [1][2][3]. Decoherence and ways to lessen the effect of decoherence on quantum states has been the subject of work [4][5][6]. Lately work has been done on experimental realization of such ideas, such as the realization of Decoherence Free Subspaces using Nickel impurities in Diamond [7]. Entanglement has been a key concept in quantum communication protocols. Quantum Repeaters and Entanglement Purification has been the subject of interest lately, as well [8][9][10].
\\
\\
For information transfer using entanglement between parties, one needs an established entangled channel-state and means of classical communication. Teleportation of an arbitrary single qubit state using a channel comprising of an EPR pair was first demonstrated by Bennett et al [11]. Lately, W-GHZ composite states have been used for Teleportation as well as Superdense Coding of arbitrary quantum states. In this paper, we have shown that the seven-qubit XWZ state can be used for teleportation of arbitrary single, double and triple qubit states. 
\\
\\
Recently, Xin-Wei Zha et al [12] arrived at a maximally entangled seven-qubit state through a numerical optimization process, following the path taken by Brown et al [13] and Borras et al [14] to find maximally entangled five-qubit and six-qubit states.

\begin{figure*}
\begin{center}
 \includegraphics[width=1\textwidth]{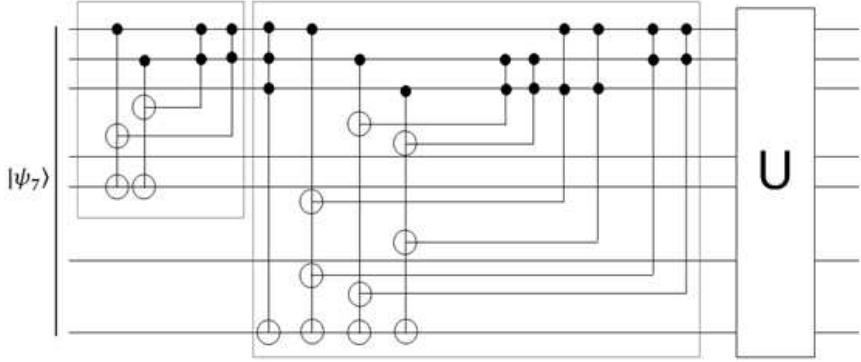}
\caption{Quantum Circuit for Physical Realization of  $|\Gamma_7\rangle$ }
\label{Figure 1: Physical Realization}       
\end{center}
\end{figure*}
\section{Quantum Teleportation of an arbitrary single qubit state}
We will consider the case where Alice possesses qubits 1, 2, 3, 4, 5, 6 and the 7th particle belongs to Bob. 
Alice wants to transport an arbitrary state $|\psi^{(1)}\rangle=\alpha|0\rangle+\beta|1\rangle$ to Bob. The combined state of the 
system is then,
\begin{equation}
|\Gamma^{(1)}_7\rangle=|\psi^{(1)}\rangle \otimes |\Gamma_7\rangle \notag
\end{equation}
Alice then measures the seven qubits in her possession via the seven qubit orthonormal states as described below:
\begin{align}
|\xi^\pm\rangle &= |0000\rangle|\Psi^0_{GHZ}\rangle-|0001\rangle|\Psi^3_{GHZ}\rangle \notag
&+|0010\rangle|\Psi^7_{GHZ}\rangle+|0011\rangle|\Psi^4_{GHZ}\rangle\\ \notag
&-|0100\rangle|\Psi^5_{GHZ}\rangle -|0101\rangle|\Psi^6_{GHZ}\rangle 
&+|0110\rangle|\Psi^2_{GHZ}\rangle+|0111\rangle|\Psi^1_{GHZ}\rangle\\ \notag 
\end{align}
\begin{align}
&\pm(|1000\rangle|\Psi^2_{GHZ}\rangle-|1001\rangle|\Psi^1_{GHZ}\rangle
-|1010\rangle|\Psi^5_{GHZ}\rangle-|1011\rangle|\Psi^6_{GHZ}\rangle\notag\\
&+|1100\rangle|\Psi^7_{GHZ}\rangle+|1101\rangle|\Psi^4_{GHZ}\rangle
+|1110\rangle|\Psi^0_{GHZ}\rangle-|1111\rangle|\Psi^3_{GHZ}\rangle)\\
\notag\\
|\nu^\pm\rangle &= |1000\rangle|\Psi^0_{GHZ}\rangle-|1001\rangle|\Psi^3_{GHZ}\rangle
+|1010\rangle|\Psi^7_{GHZ}\rangle+|1011\rangle|\Psi^4_{GHZ}\rangle\notag\\
&-|1100\rangle|\Psi^5_{GHZ}\rangle-|1101\rangle|\Psi^6_{GHZ}\rangle
+|1110\rangle|\Psi^2_{GHZ}\rangle+|1111\rangle|\Psi^1_{GHZ}\rangle\notag \\
&\pm(|0000\rangle|\Psi^2_{GHZ}\rangle-|0001\rangle|\Psi^1_{GHZ}\rangle
-|0010\rangle|\Psi^5_{GHZ}\rangle-|0011\rangle|\Psi^6_{GHZ}\rangle\notag\\
&+|0100\rangle|\Psi^7_{GHZ}\rangle+|0101\rangle|\Psi^4_{GHZ}\rangle
+|0110\rangle|\Psi^0_{GHZ}\rangle-|0111\rangle|\Psi^3_{GHZ}\rangle)
\end{align}

where 
\begin{equation}
|\Psi^{0,1}_{GHZ}\rangle = \frac{1}{\sqrt 2}[|000\rangle \pm |111\rangle]  \notag
\end{equation}
\\
\begin{equation}
|\Psi^{2,3}_{GHZ}\rangle = \frac{1}{\sqrt 2}[|001\rangle \pm |110\rangle]\notag 
\end{equation}
\\
\begin{equation}
|\Psi^{4,5}_{GHZ}\rangle = \frac{1}{\sqrt 2}[|010\rangle \pm |101\rangle]\notag 
\end{equation}
\\
\begin{equation}
|\Psi^{6,7}_{GHZ}\rangle = \frac{1}{\sqrt 2}[|100\rangle \pm |011\rangle]\notag 
\end{equation}

Alice then conveys the outcome of the measurement results to Bob via two classical bits. Bob then applies a suitable 
unitary operation from the set  {$I,\sigma_x,i\sigma_y,\sigma_z$} to recover the original state, sent by Alice. 
In this way, one can teleport an arbitrary single-qubit state using the state $|\Gamma_7\rangle$.

\section{Quantum Teleportation of an arbitrary two qubit state}

We will consider the case where Alice possesses qubits 1, 2, 3, 4 and 5, and the 6th and 7th particles belong to Bob. Alice wants to transport an arbitrary state $|\psi^{(2)}\rangle= \alpha|00\rangle + \mu|10\rangle + \gamma|01\rangle + \beta|11\rangle$ to Bob. 
The combined state of the system is then,
\begin{equation}
|\Gamma^{(2)}_7\rangle=|\psi^{(2)}\rangle \otimes |\Gamma_7\rangle
\end{equation}
\\

So, Alice prepares the combined state

\begin{align}
|\Gamma^{(2)}_7\rangle=& \alpha(A_{00}|00\rangle+A_{01}|01\rangle+A_{10}|10\rangle+A_{11}|11\rangle)\notag\\
&+\mu(B_{00}|00\rangle+B_{01}|01\rangle+B_{10}|10\rangle+B_{11}|11\rangle) \notag \\
&+\gamma(C_{00}|00\rangle+C_{01}|01\rangle+C_{10}|10\rangle+C_{11}|11\rangle) \notag \\
&+\beta(D_{00}|00\rangle+D_{01}|01\rangle+D_{10}|10\rangle+D_{11}|11\rangle)
\end{align}
\\
where\\ \\
$A_{00}=|0000\rangle|\Psi^{0}_{GHZ}\rangle+|0001\rangle|\Psi^{4}_{GHZ}\rangle-|0010\rangle|\Psi^{5}_{GHZ}\rangle-|0011\rangle|\Psi^{7}_{GHZ}\rangle$
\\
\\
$A_{11}=|0000\rangle|\Psi^{1}_{GHZ}\rangle+|0001\rangle|\Psi^{5}_{GHZ}\rangle-|0010\rangle|\Psi^{2}_{GHZ}\rangle-|0011\rangle|\Psi^{7}_{GHZ}\rangle$
\\
\\
$A_{01}=|0000\rangle|\Psi^{6}_{GHZ}\rangle-|0001\rangle|\Psi^{2}_{GHZ}\rangle+|0010\rangle|\Psi^{4}_{GHZ}\rangle+ |0011\rangle|\Psi^{0}_{GHZ}\rangle$
\\
\\
$A_{10} = -|0000\rangle|\Psi^{7}_{GHZ}\rangle-|0001\rangle|\Psi^{3}_{GHZ}\rangle+|0010\rangle|\Psi^{5}_{GHZ}\rangle+|0011\rangle |\Psi^{1}_{GHZ}\rangle$
\\
\\
$B_{00}= |1000\rangle|\Psi^{0}_{GHZ}\rangle+|1001\rangle|\Psi^{4}_{GHZ}\rangle-|1010\rangle|\Psi^{2}_{GHZ}\rangle+|1011\rangle|\Psi^{6}_{GHZ}\rangle$
\\
\\
$B_{11}=|1000\rangle|\Psi^{0}_{GHZ}\rangle+|1001\rangle|\Psi^{5}_{GHZ}\rangle-|1010\rangle|\Psi^{3}_{GHZ}\rangle-|1011\rangle|\Psi^{7}_{GHZ}\rangle$
\\
\\
$B_{01}=|1000\rangle|\Psi^{6}_{GHZ}\rangle-|1001\rangle|\Psi^{2}_{GHZ}\rangle+|1010\rangle|\Psi^{4}_{GHZ}\rangle+|1011\rangle|\Psi^{0}_{GHZ}\rangle$
\\
\\
$B_{10}=-|1000\rangle|\Psi^{7}_{GHZ}\rangle-|1001\rangle|\Psi^{3}_{GHZ}\rangle+|1010\rangle|\Psi^{5}_{GHZ}\rangle+|1011\rangle|\Psi^{1}_{GHZ}\rangle$
\\
\\
$C_{00}=|0100\rangle|\Psi^{0}_{GHZ}\rangle+|0101\rangle|\Psi^{4}_{GHZ}\rangle-|0110\rangle|\Psi^{3}_{GHZ}\rangle+|0111\rangle|\Psi^{6}_{GHZ}\rangle$
\\
\\
$C_{11}=|0100\rangle|\Psi^{1}_{GHZ}\rangle+|0101\rangle|\Psi^{5}_{GHZ}\rangle-|0110\rangle|\Psi^{3}_{GHZ}\rangle-|0111\rangle|\Psi^{7}_{GHZ}\rangle$
\\
\\
$C_{01}=|0100\rangle|\Psi^{6}_{GHZ}\rangle-|0101\rangle|\Psi^{2}_{GHZ}\rangle-|0110\rangle|\Psi^{4}_{GHZ}\rangle+|0111\rangle|\Psi^{0}_{GHZ}\rangle$
\\
\\
$C_{10}=|0100\rangle|\Psi^{7}_{GHZ}\rangle-|0101\rangle|\Psi^{3}_{GHZ}\rangle+|0110\rangle|\Psi^{5}_{GHZ}\rangle+|0111\rangle|\Psi^{1}_{GHZ}\rangle$
\\
\\
\\
$D_{00}=|1100\rangle|\Psi^{0}_{GHZ}\rangle+|1101\rangle|\Psi^{4}_{GHZ}\rangle-|1110\rangle|\Psi^{2}_{GHZ}\rangle+|1111\rangle|\Psi^{6}_{GHZ}\rangle$
\\
\\
$D_{11}=|1100\rangle|\Psi^{1}_{GHZ}\rangle+|1101\rangle|\Psi^{5}_{GHZ}\rangle-|1110\rangle|\Psi^{3}_{GHZ}\rangle-|1111\rangle|\Psi^{7}_{GHZ}\rangle$
\\
\\
$D_{01}=|1100\rangle|\Psi^{6}_{GHZ}\rangle-|1101\rangle|\Psi^{2}_{GHZ}\rangle+|1110\rangle|\Psi^{4}_{GHZ}\rangle+|1111\rangle|\Psi^{0}_{GHZ}\rangle$
\\
\\
$D_{10}=|1100\rangle|\Psi^{6}_{GHZ}\rangle-|1101\rangle|\Psi^{3}_{GHZ}\rangle+|1110\rangle|\Psi^{5}_{GHZ}\rangle+|1111\rangle|\Psi^{1}_{GHZ}\rangle$
\\
\\

where 
\begin{equation}
|\Psi^{0,1}_{GHZ}\rangle = \frac{1}{\sqrt 2}[|000\rangle \pm |111\rangle]\notag
\end{equation}

\begin{equation}
|\Psi^{2,3}_{GHZ}\rangle = \frac{1}{\sqrt 2}[|001\rangle \pm |110\rangle]\notag
\end{equation}

\begin{equation}
|\Psi^{4,5}_{GHZ}\rangle = \frac{1}{\sqrt 2}[|010\rangle \pm |101\rangle]\notag
\end{equation}

\begin{equation}
|\Psi^{6,7}_{GHZ}\rangle = \frac{1}{\sqrt 2}[|100\rangle \pm |011\rangle]\notag
\end{equation}
\begin{align}
(\alpha|00\rangle +\mu|10\rangle +\gamma|01\rangle +\beta|11\rangle )|\Gamma_7\rangle =\notag \\ 
&\frac{1}{4} \{(A_{01}+B_{11}+C_{00}+D_{01}) (\alpha|01\rangle +\mu|11\rangle +\gamma|00\rangle +\beta|01\rangle)\notag \\
&+ (A_{01}+B_{11}-C_{00}-D_{01}) (\alpha|01\rangle +\mu|11\rangle -\gamma|00\rangle -\beta|01\rangle) \notag \\
&+ (A_{01}-B_{11}+C_{00}-D_{01}) (\alpha|01\rangle -\mu|11\rangle +\gamma|00\rangle -\beta|01\rangle) \notag \\
&+ (A_{01}-B_{11}-C_{00}+D_{01}) (\alpha|01\rangle -\mu|11\rangle -\gamma|00\rangle +\beta|01\rangle) \notag \\
&+ (A_{11}+B_{01}+C_{10}+D_{00}) (\alpha|11\rangle+\mu|01\rangle+\gamma|10\rangle+\beta|00\rangle) \notag \\
&+ (A_{11}-B_{01}+C_{10}-D_{00}) (\alpha|11\rangle-\mu|01\rangle+\gamma|10\rangle-\beta|00\rangle) \notag \\
&+ (A_{11}+B_{01}-C_{10}-D_{00}) (\alpha|11\rangle+\mu|01\rangle-\gamma|10\rangle-\beta|00\rangle) \notag \\
&+ (A_{11}-B_{01}-C_{10}+D_{00}) (\alpha|11\rangle-\mu|01\rangle-\gamma|10\rangle+\beta|00\rangle) \notag \\
&+ (A_{00}+B_{10}+C_{01}+D_{11}) (\alpha|00\rangle+\mu|10\rangle+\gamma|01\rangle+\beta|11\rangle)\notag \\
&+ (A_{00}-B_{10}+C_{01}-D_{11}) (\alpha|00\rangle-\mu|10\rangle+\gamma|01\rangle-\beta|11\rangle) \notag \\
&+ (A_{00}+B_{10}-C_{01}-D_{11}) (\alpha|00\rangle+\mu|10\rangle-\gamma|01\rangle-\beta|11\rangle) \notag \\
&+ (A_{00}-B_{10}-C_{01}+D_{11}) (\alpha|00\rangle-\mu|10\rangle-\gamma|01\rangle+\beta|11\rangle) \notag \\
&+ (A_{00}+B_{10}+C_{01}+D_{11}) (\alpha|10\rangle+\mu|11\rangle+\gamma|00\rangle+\beta|01\rangle)\notag \\
&+ (A_{00}-B_{10}+C_{01}-D_{11})(\alpha|10\rangle-\mu|11\rangle+\gamma|00\rangle-\beta|01\rangle)\notag \\
&+ (A_{00}+B_{10}-C_{01}-D_{11}) (\alpha|10\rangle+\mu|11\rangle-\gamma|00\rangle-\beta|01\rangle) \notag \\
&+ (A_{00}-B_{10}-C_{01}+D_{11}) (\alpha|10\rangle-\mu|11\rangle-\gamma|00\rangle+\beta|01\rangle)\} 
\end{align}
\\
Now, Bob can carry out a combination of unitary operations, according to the given table, to obtain the original state teleported by Alice.
\\
\begin{center}
\begin{tabular}{| c | c |}
\hline
State Obtained by Bob & Unitary Operation\\ \hline
$\alpha|01\rangle+\mu|11\rangle+\gamma|00\rangle+\beta|01\rangle$  &	$ I\otimes\sigma_1$ \\ \hline
$\alpha|01\rangle+\mu|11\rangle-\gamma|00\rangle-\beta|01\rangle$ & $\sigma_3\otimes\sigma_1$ \\ \hline
$\alpha|01\rangle-\mu|11\rangle+\gamma|00\rangle-\beta|01\rangle$ & $I\otimes i\sigma_2$ \\ \hline
$\alpha|01\rangle-\mu|11\rangle-\gamma|00\rangle+\beta|01\rangle$ & $\sigma_3\otimes i\sigma_2$ \\ \hline
$\alpha|11\rangle+\mu|01\rangle+\gamma|10\rangle+\beta|00\rangle$ & $\sigma_1\otimes\sigma_1$ \\ \hline
$\alpha|11\rangle-\mu|01\rangle+\gamma|10\rangle-\beta|00\rangle$ & $\sigma_1\otimes i\sigma_2$ \\ \hline
$\alpha|11\rangle+\mu|01\rangle-\gamma|10\rangle-\beta|00\rangle$ & $i\sigma_2\otimes\sigma_1$ \\ \hline
$\alpha|11\rangle-\mu|01\rangle-\gamma|10\rangle+\beta|00\rangle$ & $i\sigma_2\otimes i\sigma_2$ \\ \hline
$\alpha|00\rangle+\mu|10\rangle+\gamma|01\rangle+\beta|11\rangle$ & $I\otimes I$ \\ \hline
$\alpha|00\rangle-\mu|10\rangle+\gamma|01\rangle-\beta|11\rangle$ & $I\otimes\sigma_3$  \\  \hline
$\alpha|00\rangle+\mu|10\rangle-\gamma|01\rangle-\beta|11\rangle$ & $\sigma_3\otimes I$ \\  \hline
$\alpha|00\rangle-\mu|10\rangle-\gamma|01\rangle+\beta|11\rangle$ & $\sigma_3\otimes\sigma_3$  \\  \hline
$\alpha|10\rangle+\mu|11\rangle+\gamma|00\rangle+\beta|01\rangle$ & $\sigma_1\otimes I$ \\  \hline
$\alpha|10\rangle-\mu|11\rangle+\gamma|00\rangle-\beta|01\rangle$ & $\sigma_1\otimes\sigma_3$ \\ \hline
$\alpha|10\rangle+\mu|11\rangle-\gamma|00\rangle-\beta|01\rangle$ & $i\sigma_2\otimes I$ \\ \hline
$\alpha|10\rangle-\mu|11\rangle-\gamma|00\rangle+\beta|01\rangle$ & $i\sigma_2\otimes\sigma_3$ \\ \hline
\end{tabular}
\end{center}
Hence we can achieve two-qubit teleportation using the Seven-Qubit State.\\ \\
\textbf{Teleportation of a Triple Qubit State }
\\ \\
We will consider the case where Alice possesses qubits 1, 2, 3, 4 and 5, and the 6th and 7th particles belong to Bob. Alice wants to transport an arbitrary state  $|\psi^{(3)}\rangle=a|000\rangle+b|001\rangle+c|010\rangle+d|011\rangle+e|100\rangle+f|101\rangle+g|110\rangle+h|111\rangle$ to Bob. \notag
\begin{align}
|\Gamma^{(3)}_7\rangle=|\psi^{(3)}\rangle \otimes |\Gamma_7\rangle \notag
\end{align}
\begin{align}
&   = aA_{000}|000\rangle + aA_{001}|001\rangle + aA_{010}|010\rangle + aA_{011}|011\rangle  \\ \notag
&+ aA_{100}|100\rangle + aA_{101}|101\rangle + aA_{110}|110\rangle + aA_{111}|111\rangle    \\ \notag
&+ bB_{000}|000\rangle + bB_{001}|001\rangle + bB_{010}|010\rangle + bB_{011}|011\rangle  \\ \notag
&+ bB_{100}|100\rangle + bB_{101}|101\rangle + bB_{110}|110\rangle + bB_{111}|111\rangle  \\   \notag
&+ cC_{000}|000\rangle + cC_{001}|001\rangle + cC_{010}|010\rangle + cC_{011}|011\rangle  \\ \notag
&+ cC_{100}|100\rangle + cC_{101}|101\rangle + cC_{110}|110\rangle + cC_{111}|111\rangle   \\  \notag
&+ dD_{000}|000\rangle + dD_{001}|001\rangle + dD_{010}|010\rangle + dD_{011}|011\rangle  \\ \notag
&+ dD_{100}|100\rangle + dD_{101}|101\rangle + dD_{110}|110\rangle + dD_{111}|111\rangle   \\  \notag
&+ eE_{000}|000\rangle + eE_{001}|001\rangle + eE_{010}|010\rangle + eE_{011}|011\rangle  \\ \notag
&+ eE_{100}|100\rangle + eE_{101}|101\rangle + eE_{110}|110\rangle + eE_{111}|111\rangle   \\  \notag
\end{align}
\begin{align}
&+ fF_{000}|000\rangle + fF_{001}|001\rangle + fF_{010}|010\rangle + fF_{011}|011\rangle  \\ \notag
&+ fF_{100}|100\rangle + fF_{101}|101\rangle + fF_{110}|110\rangle + fB_{111}|111\rangle   \\  \notag
&+ gG_{000}|000\rangle + gG_{001}|001\rangle +gG_{010}|010\rangle + gG_{011}|011\rangle  \\ \notag
&+ gG_{100}|100\rangle + gG_{101}|101\rangle + gG_{110}|110\rangle + gG_{111}|111\rangle   \\ \notag 
&+ hH_{000}|000\rangle + hH_{001}|001\rangle + hH_{010}|010\rangle + hH_{011}|011\rangle  \\ \notag
&+ hH_{100}|100\rangle + hH_{101}|101\rangle + hH_{110}|110\rangle + hH_{111}|111\rangle
\end{align}

Using the decomposition given by \underline{Appendix 1}, we have found the states and the unitary transforms that Bob has to use to convert the state he obtains to the arbitrary three-qubit state that Alice has teleported. 
\\
\\

\section{Quantum Secret Sharing}
\textbf{Proposal 1}
\\
Let us consider the situation in which Alice possesses the 1st qubit, Bob possesses qubits 2, 3, 4, 5, 6 and Charlie possesses the 7th qubit. Alice has an unknown qubit $\alpha$$\vert$0$\rangle$ + $\beta$$\vert$1$\rangle$ which she wants to share with Bob and Charlie. 
\\  \\
Now, Alice combines the unknown qubit with $\vert$$\Psi_7$ $\rangle$ and performs a Bell measurement, and conveys her outcome to Charlie by two classical bits. For instance if Alice measures in the $\vert$$\Phi_+$ $\rangle$ basis, then the Bob-Charlie system evolves into the entangled state.
\begin{align}
\alpha\vert100001\rangle-\alpha\vert000100\rangle-\alpha\vert000111\rangle-\alpha\vert001001\rangle+\alpha\vert001010\rangle+\alpha\vert010101\rangle- \notag \\
\alpha\vert010110\rangle-\alpha\vert011000\rangle+\alpha\vert011011\rangle+\alpha\vert100010\rangle+\alpha\vert101100\rangle+\alpha\vert101111\rangle-\notag \\
\alpha\vert110011\rangle+\alpha\vert111101\rangle-\alpha\vert111110\rangle+\beta\vert000000\rangle+\beta\vert000011\rangle+\beta\vert001101\rangle+\notag \\
\beta\vert001110\rangle+\beta\vert010001\rangle-\beta\vert010010\rangle+\beta\vert011100\rangle-\beta\vert011111\rangle-\beta\vert100101\rangle-\notag \\
\beta\vert000000\rangle-\beta\vert100110\rangle+\beta\vert101000\rangle+\beta\vert101011\rangle+\beta\vert110100\rangle-\beta\vert110111\rangle-\notag \\
\beta\vert111001\rangle+\beta\vert111010\rangle \notag
\end{align}
\\
Now, Bob can perform a five-qubit measurement and convey his outcome to Charlie through a classical channel. Having known the outcome of both their measurement, Charlie will obtain a certain single qubit quantum state. The outcome of the measurement performed by Bob and the state obtained by Charlie is given in the following table:

\begin{center}
\begin{longtable}{| p{5 cm} |  p{5 cm} |}
\hline
Outcome of the Measurement by Bob & State Obtained by Charlie \\ \hline
$\vert$ $A_{\pm}$ $\rangle$ & $\alpha$$\vert$0$\rangle$$\pm$$\beta$$\vert$1$\rangle$ \\ \hline
$\vert$ $B_{\pm}$ $\rangle$ & $\beta$$\vert$0$\rangle$$\pm$$\alpha$$\vert$1$\rangle$ \\ \hline

\end{longtable}
\end{center}

where \\ \\
\begin{align}
\vert A_{\pm} \rangle \\ \notag
& = -\vert00010\rangle+\vert00101\rangle-\vert01011\rangle-\vert01100\rangle \\ \notag
& +\vert10001\rangle+\vert10110\rangle-\vert11111\rangle \pm (\vert00001\rangle \\ \notag
& +\vert00110\rangle + \vert01000\rangle-\vert01111\rangle-\vert10010\rangle \\ \notag
& +\vert10101\rangle-\vert11011\rangle-\vert11100\rangle) \notag
\end{align}

\begin{align}
\vert B_{\pm} \rangle \\ \notag
&= \pm(\vert10000\rangle-\vert00011\rangle-\vert00100\rangle+\vert01010\rangle \\ \notag
& +\vert01101\rangle+\vert10111\rangle-\vert11001\rangle+\vert11110\rangle) \\ \notag
&+ \vert00000\rangle+\vert00111\rangle -\vert01001\rangle+\vert01110\rangle \\ \notag
&-\vert00000\rangle-\vert10011\rangle+\vert10100\rangle+\vert11010\rangle \\ \notag
& +\vert11101\rangle \\ \notag
\end{align}

\textbf{Proposal 2}
\\
\\
Let us consider the situation in which Alice possesses the qubits 1 and 2, Bob possesses qubits 3, 4, 5 and 6 and Charlie possesses the 7th qubit. Alice has an unknown  qubit $\alpha$$\vert$0$\rangle$ + $\beta$$\vert$1$\rangle$ which she wants to share with Bob and Charlie. 
\\
\\
Now Alice can measure in a particular basis. Suppose she measures in the GHZ Basis. Now, Bob can perform a four-qubit measurement and convey his outcome to Charlie through a classical channel. Having known the outcome of both their measurement, Charlie will obtain a certain single qubit quantum state. The outcome of the measurement performed by Bob and the state obtained by Charlie is given in the following table:

\begin{center}
\begin{longtable}{| p{5 cm} |  p{5 cm} |}
\hline
Outcome of the Measurement by Bob &	State Obtained by Charlie \\ \hline
$\vert$ $X_{\pm}$$\rangle$ & $\alpha$$\vert$0$\rangle$$\pm$$\beta$$\vert$1$\rangle$ \\ \hline
$\vert$ $Y_{\pm}$$\rangle$ & $\beta$$\vert$0$\rangle$$\pm$$\alpha$$\vert$1$\rangle$ \\ \hline
\end{longtable}
\end{center}

where \\ \\
$\vert$ $X_{\pm}$$\rangle$ = $\frac{1}{4}$ {α$\vert$0000$\rangle$+α$\vert$0111$\rangle$+α$\vert$1001$\rangle$+α$\vert$1110$\rangle$ $\pm$ (β$\vert$1001$\rangle$+β$\vert$0000$\rangle$+β$\vert$1110$\rangle$-β$\vert$0111$\rangle$)}\\ \\
$\vert$ $Y_{\pm}$$\rangle$ = $\frac{1}{4}$ {α$\vert$0001$\rangle$+α$\vert$0110$\rangle$+α$\vert$1000$\rangle$-α$\vert$1111$\rangle$ $\pm$ (β$\vert$1000$\rangle$+β$\vert$0001$\rangle$-β$\vert$1111$\rangle$-β$\vert$0110$\rangle$)}	
\\ \\ \\
\textbf{Proposal 3}
\\
Let us consider the situation in which Alice possesses the qubits 1, 2, 3 and 4, Bob possesses qubits 5 and 6 and Charlie possesses the 7th qubit. Alice has an unknown qubit $\alpha$$\vert$0$\rangle$ + $\beta$$\vert$1$\rangle$ which she wants to share with Bob and Charlie. 
\\

\begin{center}
\begin{longtable}{| p{5 cm} |  p{5 cm} |}
\hline
Outcome of Measurement by Alice & State Obtained by Bob and Charlie \\ \hline
$\vert$$A_1$$\rangle$ & $\vert$$BC_1$$\rangle$ \\ \hline
$\vert$$A_2$$\rangle$ & $\vert$$BC_2$$\rangle$ \\ \hline
$\vert$$A_3$$\rangle$ & $\vert$$BC_3$$\rangle$ \\ \hline
$\vert$$A_4$$\rangle$ & $\vert$$BC_4$$\rangle$ \\ \hline
$\vert$$A_5$$\rangle$ & $\vert$$BC_5$$\rangle$ \\ \hline
$\vert$$A_6$$\rangle$ & $\vert$$BC_6$$\rangle$ \\ \hline
$\vert$$A_7$$\rangle$ & $\vert$$BC_7$$\rangle$ \\ \hline
$\vert$$A_8$$\rangle$ & $\vert$$BC_8$$\rangle$ \\ \hline
\end{longtable}
\end{center}

where \\ \\
$\vert$$A_1$$\rangle$ = $\frac{1}{4}$($\vert$01111$\rangle$- $\vert$01011$\rangle$+ $\vert$10010$\rangle$+ $\vert$11001$\rangle$+ $\vert$11100$\rangle$+ $\vert$11101$\rangle$- $\vert$11000$\rangle$)\\
$\vert$$A_2$$\rangle$ = $\frac{1}{4}$($\vert$01111$\rangle$+ $\vert$01011$\rangle$- $\vert$10010$\rangle$- $\vert$11001$\rangle$- $\vert$11100$\rangle$+ $\vert$11101$\rangle$- $\vert$11000$\rangle$)\\
$\vert$$A_3$$\rangle$ = $\frac{1}{4}$($\vert$01111$\rangle$+ $\vert$01011$\rangle$+ $\vert$10010$\rangle$+ $\vert$11001$\rangle$+ $\vert$11100$\rangle$- $\vert$11101$\rangle$+ $\vert$11000$\rangle$)\\
$\vert$$A_4$$\rangle$ = $\frac{1}{4}$($\vert$01111$\rangle$- $\vert$01011$\rangle$- $\vert$10010$\rangle$- $\vert$11001$\rangle$- $\vert$11100$\rangle$- $\vert$11101$\rangle$+ $\vert$11000$\rangle$)\\
$\vert$$A_5$$\rangle$ = $\frac{1}{4}$($\vert$11111$\rangle$- $\vert$11011$\rangle$+ $\vert$00010$\rangle$+ $\vert$01001$\rangle$+ $\vert$01100$\rangle$+ $\vert$01101$\rangle$- $\vert$01000$\rangle$)\\
$\vert$$A_6$$\rangle$ = $\frac{1}{4}$($\vert$11111$\rangle$+ $\vert$11011$\rangle$- $\vert$00010$\rangle$- $\vert$01001$\rangle$- $\vert$01100$\rangle$+ $\vert$01101$\rangle$- $\vert$01000$\rangle$)\\
$\vert$$A_7$$\rangle$ = $\frac{1}{4}$($\vert$11111$\rangle$+ $\vert$11011$\rangle$+ $\vert$00010$\rangle$+ $\vert$01001$\rangle$+ $\vert$01100$\rangle$- $\vert$01101$\rangle$+ $\vert$01000$\rangle$)\\
$\vert$$A_8$$\rangle$ = $\frac{1}{4}$($\vert$11111$\rangle$- $\vert$11011$\rangle$- $\vert$00010$\rangle$- $\vert$01001$\rangle$- $\vert$01100$\rangle$- $\vert$01101$\rangle$+ $\vert$01000$\rangle$)\\
\\
and
\\
\\
 $\vert$$BC_1$$\rangle$ = $\alpha$$\vert$1$\rangle$$\vert$$\Phi_-$$\rangle$ + $\alpha$$\vert$0$\rangle$$\vert$$\Psi_-$$\rangle$ + $\beta$$\vert$0$\rangle$$\vert$$\Phi_+$$\rangle$ + $\beta$$\vert$1$\rangle$$\vert$$\Psi_+$$\rangle$ \\
 $\vert$$BC_2$$\rangle$ = $\alpha$$\vert$1$\rangle$$\vert$$\Phi_-$$\rangle$ - $\alpha$$\vert$0$\rangle$$\vert$$\Psi_-$$\rangle$ - $\beta$$\vert$0$\rangle$$\vert$$\Phi_+$$\rangle$ + $\beta$$\vert$1$\rangle$$\vert$$\Psi_+$$\rangle$ \\
 $\vert$$BC_3$$\rangle$ = $\alpha$$\vert$1$\rangle$$\vert$$\Phi_-$$\rangle$ - $\alpha$$\vert$0$\rangle$$\vert$$\Psi_-$$\rangle$ + $\beta$$\vert$0$\rangle$$\vert$$\Phi_+$$\rangle$ - $\beta$$\vert$1$\rangle$$\vert$$\Psi_+$$\rangle$ \\
 $\vert$$BC_4$$\rangle$ = $\alpha$$\vert$1$\rangle$$\vert$$\Phi_-$$\rangle$ + $\alpha$$\vert$0$\rangle$$\vert$$\Psi_-$$\rangle$ - $\beta$$\vert$0$\rangle$$\vert$$\Phi_+$$\rangle$ - $\beta$$\vert$1$\rangle$$\vert$$\Psi_+$$\rangle$ \\
 $\vert$$BC_5$$\rangle$ = $\beta$$\vert$1$\rangle$$\vert$$\Phi_-$$\rangle$ + $\beta$$\vert$0$\rangle$$\vert$$\Psi_-$$\rangle$ + $\alpha$$\vert$0$\rangle$$\vert$$\Phi_+$$\rangle$ + $\alpha$$\vert$1$\rangle$$\vert$$\Psi_+$$\rangle$ \\
 $\vert$$BC_6$$\rangle$ = $\beta$$\vert$1$\rangle$$\vert$$\Phi_-$$\rangle$ - $\beta$$\vert$0$\rangle$$\vert$$\Psi_-$$\rangle$ - $\alpha$$\vert$0$\rangle$$\vert$$\Phi_+$$\rangle$ + $\alpha$$\vert$1$\rangle$$\vert$$\Psi_+$$\rangle$ \\
 $\vert$$BC_7$$\rangle$ = $\beta$$\vert$1$\rangle$$\vert$$\Phi_-$$\rangle$ - $\beta$$\vert$0$\rangle$$\vert$$\Psi_-$$\rangle$ + $\alpha$$\vert$0$\rangle$$\vert$$\Phi_+$$\rangle$ - $\alpha$$\vert$1$\rangle$$\vert$$\Psi_+$$\rangle$ \\
 $\vert$$BC_8$$\rangle$ = $\beta$$\vert$1$\rangle$$\vert$$\Phi_-$$\rangle$ + $\beta$$\vert$0$\rangle$$\vert$$\Psi_-$$\rangle$ - $\alpha$$\vert$0$\rangle$$\vert$$\Phi_+$$\rangle$ - $\alpha$$\vert$1$\rangle$$\vert$$\Psi_+$$\rangle$ \\
\\
Bob can now perform a Bell measurement on his particles, and Charlie can obtain a particular resultant state by applying the appropriate unitary operation.
\\
For example, if the joint-state obtained by Bob and Charlie is $\beta$$\vert$1$\rangle$$\vert$$\Phi_-$$\rangle$+$\beta$ $\vert$0$\rangle$$\vert$$\Psi_-$$\rangle$-$\alpha$ $\vert$0$\rangle$$\vert$$\Phi_+$$\rangle$-$\alpha$$\vert$1$\rangle$$\vert$$\Psi_+$$\rangle$, one can see that Charlie will obtain the tabulated state(s) when Bob measures his qubits.
\\
\begin{center}
\begin{longtable}{| p{5 cm} |  p{5 cm} |}
\hline
Outcome of Bob’s Measurement & State Obtained by Charlie \\ \hline
$\vert$$B_1$$\rangle$ & $\vert$$C_1$$\rangle$ \\ \hline
$\vert$$B_2$$\rangle$ & $\vert$$C_2$$\rangle$ \\ \hline
$\vert$$B_3$$\rangle$ & $\vert$$C_3$$\rangle$ \\ \hline
$\vert$$B_4$$\rangle$ & $\vert$$C_4$$\rangle$ \\ \hline
\end{longtable}
\end{center}

where
\\
$\vert$$B_1$$\rangle$ =  $\frac{1}{\sqrt{2}}$$\vert$01$\rangle$,
$\vert$$B_2$$\rangle$ =  $\frac{1}{\sqrt{2}}$$\vert$10$\rangle$, 
$\vert$$B_3$$\rangle$ =  $\frac{1}{\sqrt{2}}$$\vert$11$\rangle$, 
$\vert$$B_4$$\rangle$ =  $\frac{1}{\sqrt{2}}$$\vert$00$\rangle$ \\
\\
and \\ \\
$\vert$$C_1$$\rangle$ = $\alpha$$\vert$0$\rangle$ + $\beta$ $\vert$1$\rangle$,
$\vert$$C_2$$\rangle$ = $\alpha$$\vert$0$\rangle$ - $\beta$ $\vert$1$\rangle$ \\
$\vert$$C_3$$\rangle$ = $\alpha$$\vert$1$\rangle$ + $\beta$ $\vert$0$\rangle$, 
$\vert$$C_4$$\rangle$ = $\alpha$$\vert$1$\rangle$ - $\beta$ $\vert$0$\rangle$ \\

\textbf{Conclusion}
\\
\\
We have shown that the new seven-qubit state founded by Xin-Wei Zha et al. has many useful applications in quantum information and computation. We use the XWZ state for perfect teleportation of arbitrary one, two and three qubit states, and Quantum state sharing of arbitrary one qubit states under different scenarios and proposals. This state is also a very useful resource for superdense coding. Decoherence property of this state needs careful investigation in case of practical applications. 
\\
\\

\textbf{Acknowledgement}
\\
\\
This work was supported by the National Initiative for Undergraduate Science (NIUS) program undertaken by the Homi Bhabha Centre for Science Education (HBCSE - TIFR), Mumbai, India. The authors acknowledge Prof. V. Singh of HBCSE for being a part of this effort.
\\
\\
\textbf{References}
\\

\lbrack 1\rbrack Saha, Debashis, and Prasanta K. Panigrahi. "N-qubit quantum teleportation, information splitting and superdense coding through the composite GHZ–Bell channel." Quantum Information Processing 11.2 (2012): 615-628.
\\
\\
\lbrack 2\rbrack Zhao, Ming-Jing, et al. "Multiqubit quantum teleportation." arXiv preprint arXiv:1209.4538 (2012).
\\
\\
\lbrack 3\rbrack Zhang, Zhi-hua, Lan Shu, and Zhi-wen Mo. "Quantum teleportation and superdense coding through the composite W-Bell channel." Quantum Information Processing (2012): 1-11.
\\
\\
\lbrack 4\rbrack Wang, Xiaoting, Mark Byrd, and Kurt Jacobs. "Numerical method for finding decoherence-free subspaces and its applications." Physical Review A 87.1 (2013): 012338.
\\
\\
\lbrack 5\rbrack van der Sar, T., et al. "Decoherence-protected quantum gates for a hybrid solid-state spin register." Nature 484.7392 (2012): 82-86.
\\
\\
\lbrack 6\rbrack Nakahara, Mikio. "Frank Gaitan: Quantum error correction and fault tolerant quantum computing." Quantum Information Processing 11.2 (2012): 629-631.
\\
\\
\lbrack 7\rbrack Chanier, Thomas, C. E. Pryor, and Michael E. Flatte. "Substitutional nickel impurities in diamond: Decoherence-free subspaces for quantum information processing." EPL (Europhysics Letters) 99.6 (2012): 67006.
\\
\\
\lbrack 8\rbrack Tittel, Wolfgang. "How to Overcome the Distance Barrier in Quantum Communication: Quantum Repeaters and Quantum Memory." CLEO: Science and Innovations. Optical Society of America, 2012.
\\
\\
\lbrack 9\rbrack Zwerger, M., W. Dür, and H. J. Briegel. "Measurement-based quantum repeaters." Physical Review A 85.6 (2012): 062326.
\\
\\
\lbrack 10\rbrack Brion, Etienne, et al. "Quantum repeater with Rydberg-blocked atomic ensembles in fiber-coupled cavities." Physical Review A 85.4 (2012): 042324.
\\
\\
\lbrack 11\rbrack C. H. Bennett, G. Brassard, C. Crepeau, R. Jozsa, A. Peres, and W. K. Wootters, Phys. Rev. Lett. 70, 1895 (1993).
\\
\\
\lbrack 12\rbrack Zha, Xin-Wei, et al. "A genuine maximally seven-qubit entangled state." arXiv preprint arXiv:1110.5011 (2011).
\\
\\
\lbrack 13\rbrack Brown, Iain DK, et al. "Searching for highly entangled multi-qubit states."Journal of Physics A: Mathematical and General 38.5 (2005): 1119.
\\
\\
\lbrack 14\rbrack Borras, A., et al. "Multiqubit systems: highly entangled states and entanglement distribution." Journal of Physics A: Mathematical and Theoretical40.44 (2007): 13407.
\\
\\
\lbrack 15\rbrack Dan Liu, Xin Zhao, and Gui Lu Long. “Multiple Entropy Measures for Multipartite Quantum Entanglement”, arXiv:0705.3904v4 [quant-ph] 22 Jun 2007
\clearpage

\title{Appendix}

\textbf{Appendix 1}
\\
\\
 $\vert$ $A_{000}$$\rangle$ = $\vert$ 0000000$\rangle$ + $\vert$ 0000101$\rangle$ - $\vert$ 0001011$\rangle$ + $\vert$ 0001110$\rangle$ \\
 $\vert$ $A_{001}$$\rangle$ = $\vert$ 0000010$\rangle$ + $\vert$ 0001001$\rangle$ + $\vert$ 0001100$\rangle$ - $\vert$ 0000111$\rangle$ \\
 $\vert$ $A_{010}$$\rangle$ = $\vert$ 0000111$\rangle$ - $\vert$ 0000010$\rangle$ + $\vert$ 0001001$\rangle$ + $\vert$ 0001100$\rangle$ \\
 $\vert$ $A_{011}$$\rangle$ = $\vert$ 0000000$\rangle$ + $\vert$ 0000101$\rangle$ + $\vert$ 0001011$\rangle$ - $\vert$ 0001110$\rangle$ \\
 $\vert$ $A_{100}$$\rangle$ = $\vert$ 0000011$\rangle$ + $\vert$ 0000110$\rangle$ - $\vert$ 0001000$\rangle$ + $\vert$ 0001101$\rangle$ \\
 $\vert$ $A_{101}$$\rangle$ = $\vert$ 0000001$\rangle$ - $\vert$ 0000100$\rangle$ + $\vert$ 0001010$\rangle$ + $\vert$ 0001111$\rangle$ \\
 $\vert$ $A_{110}$$\rangle$ = $\vert$ 0000001$\rangle$ - $\vert$ 0000100$\rangle$ - $\vert$ 0001010$\rangle$ - $\vert$ 0001111$\rangle$ \\
 $\vert$ $A_{111}$$\rangle$ = $\vert$ 0001101$\rangle$ - $\vert$ 0000011$\rangle$ - $\vert$ 0000110$\rangle$ - $\vert$ 0001000$\rangle$ \\ \\

 $\vert$ $B_{000}$$\rangle$ = $\vert$ 0010000$\rangle$ + $\vert$ 0010101$\rangle$ - $\vert$ 0011011$\rangle$ + $\vert$ 0011110$\rangle$ \\
 $\vert$ $B_{001}$$\rangle$ = $\vert$ 0010010$\rangle$ + $\vert$ 0011001$\rangle$  + $\vert$ 0011100$\rangle$ - $\vert$ 0010111$\rangle$ \\
 $\vert$ $B_{010}$$\rangle$ = $\vert$ 0010111$\rangle$ - $\vert$ 0010010$\rangle$ + $\vert$ 0011001$\rangle$ + $\vert$ 0011100$\rangle$ \\
 $\vert$ $B_{011}$$\rangle$ = $\vert$ 0010000$\rangle$ + $\vert$ 0010101$\rangle$+ $\vert$ 0011011$\rangle$- $\vert$ 0011110$\rangle$ \\
 $\vert$ $B_{100}$$\rangle$ = $\vert$ 0010011$\rangle$ + $\vert$ 0010110$\rangle$ - $\vert$ 0011000$\rangle$ + $\vert$ 0011101$\rangle$ \\
 $\vert$ $B_{101}$$\rangle$ = $\vert$ 0010001$\rangle$ - $\vert$ 0010100$\rangle$ + $\vert$ 0011010$\rangle$ + $\vert$ 0011111$\rangle$ \\
 $\vert$ $B_{110}$$\rangle$ = $\vert$ 0010001$\rangle$ - $\vert$ 0010100$\rangle$ - $\vert$ 0011010$\rangle$ - $\vert$ 0011111$\rangle$ \\
 $\vert$ $B_{111}$$\rangle$ = $\vert$ 0011101$\rangle$ - $\vert$ 0010011$\rangle$ - $\vert$ 0010110$\rangle$ - $\vert$ 0011000$\rangle$ \\
 \\
 $\vert$ $C_{000}$$\rangle$ = $\vert$ 0100000$\rangle$ + $\vert$ 0100101$\rangle$ - $\vert$ 0101011$\rangle$ + $\vert$ 0101110$\rangle$ \\
 $\vert$ $C_{001}$$\rangle$= $\vert$ 0100010$\rangle$ + $\vert$ 0101001$\rangle$ + $\vert$ 0101100$\rangle$ - $\vert$ 0100111$\rangle$ \\
 $\vert$ $C_{010}$$\rangle$= $\vert$ 0100111$\rangle$ - $\vert$ 0100010$\rangle$ + $\vert$ 0101001$\rangle$ + $\vert$ 0101100$\rangle$ \\
 $\vert$ $C_{011}$$\rangle$= $\vert$ 0100000$\rangle$ + $\vert$ 0100101$\rangle$ + $\vert$ 0101011$\rangle$ - $\vert$ 0101110$\rangle$ \\
 $\vert$ $C_{100}$$\rangle$= $\vert$ 0100011$\rangle$ + $\vert$ 0100110$\rangle$ - $\vert$ 0101000$\rangle$ + $\vert$ 0101101$\rangle$ \\
 $\vert$ $C_{101}$$\rangle$= $\vert$ 0100001$\rangle$ - $\vert$ 0100100$\rangle$ + $\vert$ 0101010$\rangle$ + $\vert$ 0101111$\rangle$ \\
 $\vert$ $C_{110}$$\rangle$= $\vert$ 0100001$\rangle$ - $\vert$ 0100100$\rangle$ - $\vert$ 0101010$\rangle$ - $\vert$ 0101111$\rangle$ \\
 $\vert$ $C_{111}$$\rangle$= $\vert$ 0101101$\rangle$ - $\vert$ 0100011$\rangle$ - $\vert$ 0100110$\rangle$ - $\vert$ 0101000$\rangle$ \\
 \\
 $\vert$ $D_{000}$$\rangle$= $\vert$ 0110000$\rangle$ + $\vert$ 0110101$\rangle$ - $\vert$ 0111011$\rangle$ + $\vert$ 0111110$\rangle$ \\
 $\vert$ $D_{001}$$\rangle$= $\vert$ 0110010$\rangle$ + $\vert$ 0111001$\rangle$ + $\vert$ 0111100$\rangle$ - $\vert$ 0110111$\rangle$ \\
 $\vert$ $D_{010}$$\rangle$= $\vert$ 0110111$\rangle$ - $\vert$ 0110010$\rangle$ + $\vert$ 0111001$\rangle$ + $\vert$ 0111100$\rangle$ \\
 $\vert$ $D_{011}$$\rangle$= $\vert$ 0110000$\rangle$ + $\vert$ 0110101$\rangle$ + $\vert$ 0111011$\rangle$ - $\vert$ 0111110$\rangle$ \\
 $\vert$ $D_{100}$$\rangle$= $\vert$ 0110011$\rangle$ + $\vert$ 0110110$\rangle$ - $\vert$ 0111000$\rangle$ + $\vert$ 0111101$\rangle$ \\
 $\vert$ $D_{101}$$\rangle$= $\vert$ 0110001$\rangle$ - $\vert$ 0110100$\rangle$ + $\vert$ 0111010$\rangle$ + $\vert$ 0111111$\rangle$ \\
 $\vert$ $D_{110}$$\rangle$= $\vert$ 0110001$\rangle$ - $\vert$ 0110100$\rangle$ - $\vert$ 0111010$\rangle$ - $\vert$ 0111111$\rangle$ \\
 $\vert$ $D_{111}$$\rangle$= $\vert$ 0111101$\rangle$ - $\vert$ 0110011$\rangle$ - $\vert$ 0110110$\rangle$ - $\vert$ 0111000$\rangle$ \\
 \\
 $\vert$ $E_{000}$$\rangle$= $\vert$ 1000000$\rangle$ + $\vert$ 1000101$\rangle$ - $\vert$ 1001011$\rangle$ + $\vert$ 1001110$\rangle$ \\
 $\vert$ $E_{001}$$\rangle$= $\vert$ 1000010$\rangle$ + $\vert$ 1001001$\rangle$ + $\vert$ 1001100$\rangle$ - $\vert$ 1000111$\rangle$ \\
 $\vert$ $E_{010}$$\rangle$= $\vert$ 1000111$\rangle$ - $\vert$ 1000010$\rangle$ + $\vert$ 0001001$\rangle$ + $\vert$ 0001100$\rangle$ \\
 $\vert$ $E_{011}$$\rangle$= $\vert$ 1000000$\rangle$ + $\vert$ 1000101$\rangle$ + $\vert$ 1001011$\rangle$ - $\vert$ 1001110$\rangle$ \\
 $\vert$ $E_{100}$$\rangle$= $\vert$ 1000011$\rangle$ + $\vert$ 1000110$\rangle$ - $\vert$ 1001000$\rangle$ + $\vert$ 1001101$\rangle$ \\
 $\vert$ $E_{101}$$\rangle$= $\vert$ 1000001$\rangle$ - $\vert$ 1000100$\rangle$ + $\vert$ 1001010$\rangle$ + $\vert$ 1001111$\rangle$ \\
 $\vert$ $E_{110}$$\rangle$= $\vert$ 1000001$\rangle$ - $\vert$ 1000100$\rangle$ - $\vert$ 1001010$\rangle$ - $\vert$ 1001111$\rangle$ \\
 $\vert$ $E_{111}$$\rangle$= $\vert$ 1001101$\rangle$ - $\vert$ 1000011$\rangle$ - $\vert$ 1000110$\rangle$ - $\vert$ 1001000$\rangle$ \\
 \\
 $\vert$ $F_{000}$$\rangle$= $\vert$ 1010000$\rangle$ + $\vert$ 1010101$\rangle$ - $\vert$ 1011011$\rangle$ + $\vert$ 1011110$\rangle$ \\
 $\vert$ $F_{001}$$\rangle$= $\vert$ 1010010$\rangle$ + $\vert$ 1011001$\rangle$ + $\vert$ 1011100$\rangle$ - $\vert$ 1010111$\rangle$ \\
 $\vert$ $F_{010}$$\rangle$= $\vert$ 1010111$\rangle$ - $\vert$ 1010010$\rangle$ + $\vert$ 1011001$\rangle$ + $\vert$ 1011100$\rangle$ \\
 $\vert$ $F_{011}$$\rangle$= $\vert$ 1010000$\rangle$ + $\vert$ 1010101$\rangle$ + $\vert$ 1011011$\rangle$ - $\vert$ 1011110$\rangle$ \\
 $\vert$ $F_{100}$$\rangle$= $\vert$ 1010011$\rangle$ + $\vert$ 1010110$\rangle$ - $\vert$ 1011000$\rangle$ + $\vert$ 1011101$\rangle$ \\
 $\vert$ $F_{101}$$\rangle$= $\vert$ 1010001$\rangle$ - $\vert$ 1010100$\rangle$ + $\vert$ 1011010$\rangle$ + $\vert$ 1011111$\rangle$ \\
 $\vert$ $F_{110}$$\rangle$= $\vert$ 1010001$\rangle$ - $\vert$ 1010100$\rangle$ - $\vert$ 1011010$\rangle$ - $\vert$ 1011111$\rangle$ \\
 $\vert$ $F_{111}$$\rangle$= $\vert$ 1011101$\rangle$ - $\vert$ 1010011$\rangle$ - $\vert$ 1010110$\rangle$ - $\vert$ 1011000$\rangle$ \\
 \\
 $\vert$ $G_{000}$$\rangle$= $\vert$ 1100000$\rangle$ + $\vert$ 1100101$\rangle$ - $\vert$ 1101011$\rangle$ + $\vert$ 1101110$\rangle$ \\
 $\vert$ $G_{001}$$\rangle$= $\vert$ 1100010$\rangle$ + $\vert$ 1101001$\rangle$ + $\vert$ 1101100$\rangle$ - $\vert$ 1100111$\rangle$ \\
 $\vert$ $G_{010}$$\rangle$= $\vert$ 1100111$\rangle$ - $\vert$ 1100010$\rangle$ + $\vert$ 1101001$\rangle$ + $\vert$ 1101100$\rangle$ \\
 $\vert$ $G_{011}$$\rangle$= $\vert$ 1100000$\rangle$ + $\vert$ 1100101$\rangle$ + $\vert$ 1101011$\rangle$ - $\vert$ 1101110$\rangle$ \\
 $\vert$ $G_{100}$$\rangle$= $\vert$ 1100011$\rangle$ + $\vert$ 1100110$\rangle$ - $\vert$ 1101000$\rangle$ + $\vert$ 1101101$\rangle$ \\
 $\vert$ $G_{101}$$\rangle$= $\vert$ 1100001$\rangle$ - $\vert$ 1100100$\rangle$ + $\vert$ 1101010$\rangle$ + $\vert$ 1101111$\rangle$ \\
 $\vert$ $G_{110}$$\rangle$= $\vert$ 1100001$\rangle$ - $\vert$ 1100100$\rangle$ - $\vert$ 1101010$\rangle$ - $\vert$ 1101111$\rangle$ \\
 $\vert$ $G_{111}$$\rangle$= $\vert$ 1101101$\rangle$ - $\vert$ 1100011$\rangle$ - $\vert$ 0000110$\rangle$ - $\vert$ 0001000$\rangle$ \\
 \\
 $\vert$ $H_{000}$$\rangle$= $\vert$ 1110000$\rangle$ + $\vert$ 1110101$\rangle$ - $\vert$ 1111011$\rangle$ + $\vert$ 1111110$\rangle$ \\
 $\vert$ $H_{001}$$\rangle$= $\vert$ 1110010$\rangle$ + $\vert$ 1111001$\rangle$ + $\vert$ 1111100$\rangle$ - $\vert$ 1110111$\rangle$ \\
 $\vert$ $H_{010}$$\rangle$= $\vert$ 1110111$\rangle$ - $\vert$ 1110010$\rangle$ + $\vert$ 1111001$\rangle$ + $\vert$ 1111100$\rangle$ \\
 $\vert$ $H_{011}$$\rangle$= $\vert$ 1110000$\rangle$ + $\vert$ 1110101$\rangle$ + $\vert$ 1111011$\rangle$ - $\vert$ 1111110$\rangle$ \\
 $\vert$ $H_{100}$$\rangle$= $\vert$ 1110011$\rangle$ + $\vert$ 1110110$\rangle$ - $\vert$ 1111000$\rangle$ + $\vert$ 1111101$\rangle$ \\
 $\vert$ $H_{101}$$\rangle$= $\vert$ 1110001$\rangle$ - $\vert$ 1110100$\rangle$ + $\vert$ 1111010$\rangle$ + $\vert$ 1111111$\rangle$ \\
 $\vert$ $H_{110}$$\rangle$= $\vert$ 1110001$\rangle$ - $\vert$ 1110100$\rangle$ - $\vert$ 1111010$\rangle$ - $\vert$ 1111111$\rangle$ \\
 $\vert$ $H_{111}$$\rangle$= $\vert$ 1111101$\rangle$ - $\vert$ 1110011$\rangle$ - $\vert$ 1110110$\rangle$ - $\vert$ 1111000$\rangle$ \\ \notag 
\\
\clearpage
\textbf{Appendix 2}
\\

\begin{align}
(a \vert 000\rangle + b\vert 001\rangle + c\vert 010\rangle + d\vert 011\rangle + e\vert 100\rangle + f\vert 101\rangle + g\vert 110\rangle + h\vert 111\rangle) \vert \Psi_7 \rangle  \notag
\end{align}
\begin{align}
=  \sum\limits_{permutations} ((-1)^{I_1} A_{a_1 a_2 a_3} + (-1)^{I_2} B_{b_1 b_2 b_3}  + (-1)^{I_3} C_{{c_1 c_2 c_3}} +  (-1)^{I_4} D_{{d_1 d_2 d_3}} \notag
\end{align}
\begin{align}
& + (-1)^{I_5} E_{{e_1 e_2 e_3}} + (-1)^{I_6} F_{{f_1 f_2 f_3}} + (-1)^{I_7} G_{{g_1 g_2 g_3}} + (-1)^{I_8} H_{{h_1 h_2 h_3}})  \\ \notag 
\end{align}
\begin{align}
& ((-1)^{I_1} a\vert a_{1}a_{2}a_{3}\rangle + (-1)^{I_2} b\vert b_{1}b_{2}b_{3}\rangle  \\ \notag 
& + (-1)^{I_3} c\vert c_{1}c_{2}c_{3}\rangle + (-1)^{I_4} d\vert d_{1}d_{2}d_{3}\rangle  \\ \notag 
& + (-1)^{I_5} e\vert e_{1}e_{2}e_{3}\rangle + (-1)^{I_6} f\vert f_{1}f_{2}f_{3}\rangle  \\ \notag 
& + (-1)^{I_7} g\vert g_{1}g_{2}g_{3}\rangle + (-1)^{I_8} h\vert h_{1}h_{2}h_{3}\rangle) \\
\end{align}

where $I_i$ (i=1, 2, 3, 4, 5, 6, 7, 8) can take values 0 or 1 independently, and $L_j$ (L=a, b, c, d, e, f, g, h; j = 1, 2, 3) can take values 0 or 1 independently. The summation is over all possible permutation states obtained.
\\
\\
\\
\textbf{Appendix 3}
\\
\textbf{Transformations}

$P_2$: Projection of Second Component -
\[ \left( 
\begin{array}{cc}
0 & 0 \\
0 & 1 \end{array} \right)\] 
\\

$P_1$: Projection of First Component -
\[ \left( \begin{array}{cc}
1 & 0 \\
0 & 0 \end{array} \right)\] 
\\

$\sigma_1$-
\[ \left( \begin{array}{cc}
0 & 1 \\
1 & 0 \end{array} \right)\] 
\\

$\sigma_2$ -
\[ \left( \begin{array}{cc}
0 & i \\
-i & 0 \end{array} \right)\] 
\\

$\sigma_3$ -
\[ \left( \begin{array}{cc}
1 & 0 \\
0 & -1 \end{array} \right)\] 
\\

$F_2$: Flip and Projection of Second Component - 
\[ \left( \begin{array}{cc}
0 & 0 \\
1 & 0 \end{array} \right)\] 
\\

$F_1$: Flip and Projection of First Component - 
\[ \left( \begin{array}{cc}
0 & 1 \\
0 & 0 \end{array} \right)\] 
\\

\begin{center}
\begin{longtable}{|  p {7 cm} |  p {7 cm} | }
\hline
State Obtained by Bob & Short-Hand Form \\ \hline
a$\vert$ 000$\rangle$ + b$\vert$ 001$\rangle$ + c$\vert$ 010$\rangle$ + d$\vert$ 011$\rangle$ + e$\vert$ 100$\rangle$ + f$\vert$ 101$\rangle$ + g$\vert$ 110$\rangle$ + h$\vert$ 111$\rangle$ & $I_2$$\otimes$$I_2$$\otimes$$I_2$ \\ \hline
a$\vert$ 000$\rangle$ - b$\vert$ 001$\rangle$ + c$\vert$ 010$\rangle$ + d$\vert$ 011$\rangle$ + e$\vert$ 100$\rangle$ + f$\vert$ 101$\rangle$ + g$\vert$ 110$\rangle$ + h$\vert$ 111$\rangle$ & $I_2$$\otimes$$I_2$$\otimes$$P_2$ + $\sigma_3$$\otimes$$P_1$$\otimes$$P_1$ + $I_2$$\otimes$$P_2$$\otimes$$P_1$ \\ \hline
a$\vert$ 000$\rangle$ + b$\vert$ 001$\rangle$ + c$\vert$ 010$\rangle$ - d$\vert$ 011$\rangle$ + e$\vert$ 100$\rangle$ + f$\vert$ 101$\rangle$ + g$\vert$ 110$\rangle$ + h$\vert$ 111$\rangle$ & $I_2$$\otimes$$I_2$$\otimes$$P_2$ + $I_2$$\otimes$$P_1$$\otimes$$P_1$ + $\sigma_3$$\otimes$$P_2$$\otimes$$P_1$ \\ \hline
a$\vert$ 000$\rangle$ + b$\vert$ 001$\rangle$ + c$\vert$ 010$\rangle$ + d$\vert$ 011$\rangle$ + e$\vert$ 100$\rangle$ - f$\vert$ 101$\rangle$ + g$\vert$ 110$\rangle$ + h$\vert$ 111$\rangle$ & $I_2$$\otimes$$I_2$$\otimes$$P_1$ + $\sigma_3$$\otimes$$P_1$$\otimes$$P_2$ + $I_2$$\otimes$$P_2$$\otimes$$P_2$ \\ \hline
a$\vert$ 000$\rangle$ + b$\vert$ 001$\rangle$ + c$\vert$ 010$\rangle$ + d$\vert$ 011$\rangle$ + e$\vert$ 100$\rangle$ + f$\vert$ 101$\rangle$ + g$\vert$ 110$\rangle$ - h$\vert$ 111$\rangle$ & $I_2$$\otimes$$I_2$$\otimes$$P_1$ + $\sigma_3$$\otimes$$P_2$$\otimes$$P_2$ + $I_2$$\otimes$$P_2$$\otimes$$P_2$ \\ \hline
a$\vert$ 000$\rangle$ - b$\vert$ 001$\rangle$ + c$\vert$ 010$\rangle$ - d$\vert$ 011$\rangle$ + e$\vert$ 100$\rangle$ + f$\vert$ 101$\rangle$ + g$\vert$ 110$\rangle$ + h$\vert$ 111$\rangle$ & $I_2$$\otimes$$I_2$$\otimes$$P_2$ + $\sigma_3$$\otimes$$P_1$$\otimes$$P_1$ + $\sigma_3$$\otimes$$P_2$$\otimes$$P_1$ \\
\end{longtable}
\end{center}
\clearpage

\end{document}